\documentclass[runningheads]{llncs}
\usepackage{graphicx}
\usepackage[utf8]{inputenc}
\usepackage{amssymb}
\usepackage{hyperref}
\usepackage{mathtools}
\usepackage{amsmath}
\usepackage{enumerate}
\usepackage{tikz}
\usepackage{amsmath}
\usepackage{amssymb}
\usepackage{algorithm}
\usepackage[noend]{algpseudocode}
\usepackage{multirow}
\usepackage{enumitem}
\usepackage{array}
\usepackage{comment}
\usepackage{subcaption}
\usepackage{caption}
\usepackage{float}
\usepackage{booktabs}

\begin{document}
\title{Improving Medical Image Classification in Noisy Labels Using Only Self-supervised Pretraining}

\authorrunning{B. Khanal et al.}

\author{Bidur Khanal\inst{1}\and
Binod Bhattarai\inst{4}\and
Bishesh Khanal\inst{3}\and
Cristian A. Linte\inst{1,2}}
\titlerunning{Improving Medical Image Classification in Noisy Labels}

\institute{ Center for Imaging Science, RIT, Rochester, NY, USA \and Biomedical Engineering, RIT, Rochester, NY, USA \and NepAl Applied Mathematics and Informatics Institute for Research (NAAMII) \and University of Aberdeen, Aberdeen, UK\\}
%
\maketitle              
\begin{abstract}

Noisy labels hurt deep learning-based supervised image classification performance as the models may overfit the noise and learn corrupted feature extractors.
For natural image classification training with noisy labeled data, model initialization with contrastive self-supervised pretrained weights has shown to reduce feature corruption and improve classification performance.
However, no works have explored: i) how other self-supervised approaches, such as pretext task-based pretraining, impact the learning with noisy label, and ii) any self-supervised pretraining methods alone for medical images in noisy label settings.
Medical images often feature smaller datasets and subtle inter-class variations, requiring human expertise to ensure correct classification. Thus, it is not clear if the methods improving learning with noisy labels in natural image datasets such as CIFAR would also help with medical images.
In this work, we explore contrastive and pretext task-based self-supervised pretraining to initialize the weights of a deep learning classification model for two medical datasets with self-induced noisy labels---\textit{NCT-CRC-HE-100K} tissue histological images and \textit{COVID-QU-Ex} chest X-ray images.
Our results show that models initialized with pretrained weights obtained from self-supervised learning can effectively learn better features and improve robustness against noisy labels.

\keywords{medical image classification \and label noise \and learning with noisy labels \and self-supervised pretraining \and warm-up obstacle \and feature extraction.}
\end{abstract}
\section{Introduction}
Medical image classification using supervised learning relies on large amounts of representative data with accurately annotated labels to achieve good generalization.
However, recent practices of crowd-sourcing for data labeling or automatically generating labels from patients' medical reports using algorithms, and the high variability among expert annotators introduce higher levels of label noise in medical datasets. Moreover, supervised deep learning is highly susceptible to label noise as the models can easily overfit the noisy labels, leading to corrupt representation learning and compromising generalizability \cite{lee2019robust,zhang2021understanding,khanal2021does,khanal2023investigating}.
Correcting label noise in large medical image datasets is expensive and requires extensive human resources and time-consuming protocols.
Several methods for learning with noisy labels (LNL) have been introduced in natural image datasets to minimize the influence of label noise on the training \cite{hu2019simple,chen2019understanding,han2018co,Wei_2020_CVPR,song2019does,Li2020DivideMix:}.
Similar methods, with adjustments, have also been applied to medical image classification \cite{ju2022improving,xue2022robust,liu2021co,zhou2023combating}.

Many LNL methods rely on a warm-up phase, a small number of initial epochs during which the model is trained directly using all the noisy training data \cite{han2018co,Li2020DivideMix:,liu2020early,ju2022improving}.
While the warm-up phase is important to kickstart the model and learn basic features important for proper separation of noisy labels from clean labels at a later phase \cite{zhang2021codim,zheltonozhskii2022contrast}, the high noise rate makes it challenging to avoid memorizing wrong labels and learning poor feature extractors. Zheltonozhskii et al. \cite{zheltonozhskii2022contrast} referred to this issue as the ``warm-up obstacle".
One may use supervised pretraining to learn good feature extractors and train with noisy labels to mitigate the warm-up obstacle.
However, this approach presents challenges in medical datasets due to the limited availability of large labeled datasets that closely align with the given new dataset.
Alternatively, if existing medical datasets already contain valuable metadata such as gender and age information, one may pretrain to predict such auxiliary information before proceeding with training on the main task involving noisy labels.
Such an approach could minimize feature corruption, as the auxiliary tasks are relatively straightforward and less likely to contain label noise. However, if the datasets lack such metadata, another approach is to use self-supervised learning techniques for pretraining to learn feature extractors, without relying on any labels.

\begin{figure}[h!]
\centering
\includegraphics[trim = {0 0 0.5em 0},clip, width=1\linewidth]{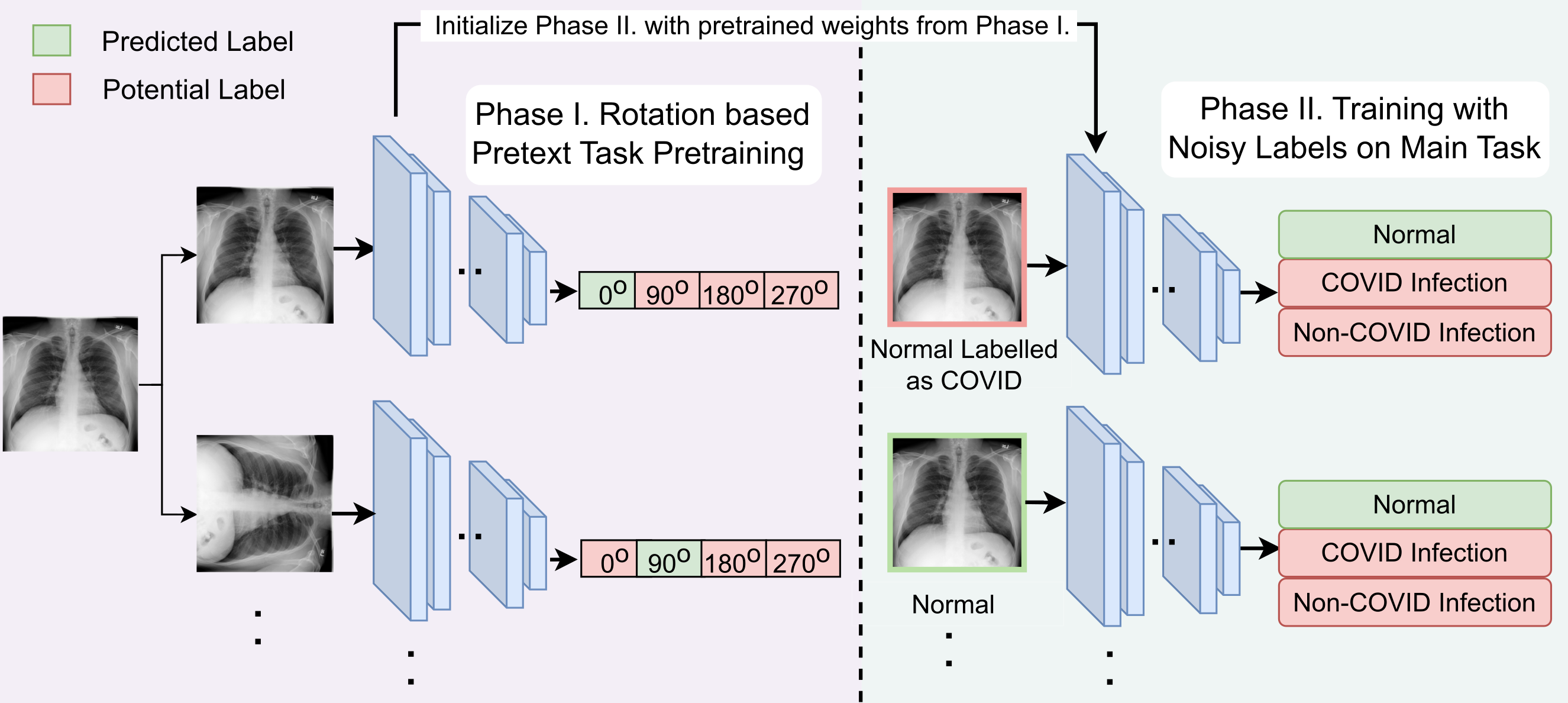}
\caption{Our approach involves two phases: I. Pretrain the model with pretext task-based self-supervised technique (left), and II. Retrain the pretrained model on medical image classification with noisy labels using LNL approaches (right).}
\label{fig:pipeline}
\end{figure}

Some studies have demonstrated the benefits of contrastive learning-based self-supervised pretraining to improve robustness against noisy labels in natural image datasets \cite{zheltonozhskii2022contrast,zhang2021codim}. However, no extensive study has been conducted to investigate which self-supervised pretraining is suitable for a specific scenario, therefore providing no such prior knowledge that can be adapted and used in medical image classification. Additionally, medical images come with some caveats that make it challenging to apply various self-supervised techniques in medical datasets (discussed in Sec \ref{self-supervision_pretext_task}).

In this work, we investigate contrastive learning and propose simple and intuitive pretext task-based self-supervised pretraining approaches to improve robustness against noisy labels in the medical image classification problem.
We show that self-supervised pretraining can significantly improve the robustness against noisy labels in the existing classification framework. Furthermore, we explored the implications of this pretraining approach on existing LNL methods by pretraining the models before the warm-up phase.

Our contributions can be summarized as follows: \textbf{1)} To our knowledge, we are the first to investigate the use of only self-supervised pretraining to improve robustness in the presence of noisy labels for medical image classification; \textbf{2)} We propose the use of pretext task-based self-supervised pretraining in classification with noisy labels, which hasn't been studied even with natural image datasets; \textbf{3)} Using two representative datasets, namely X-ray and histopathology images, induced with label noise at various rates, we show that self-supervised pretraining alone improves the feature extractor, thus helping overcome the warm-up obstacle in LNL methods, yielding significantly improved performance while reducing the label memorization.

\section{Related Works}

\subsection{Learning with Noisy Labels in Medical Images}
Several methods have been proposed to robustly train medical image classifiers with noisy labels \cite{karimi2020deep}. Pham et al. \cite{pham2021interpreting} used label smoothing to reduce the impact of noisy labels in thoracic disease classification. Dgani et al. \cite{dgani2018training} introduced a noise layer and modified the network architecture to address unreliable labels in breast classification. Le et al. \cite{le2019pancreatic} used a sample reweighting technique to robustly train a pancreatic cancer detection model with noisy labels, while Xue et al. \cite{xue2019robust} used a similar reweighting technique for skin lesion classification with noisy datasets.

Ju et al. \cite{ju2022improving} used dual-uncertainty estimation to tackle two cases: label noise due to disagreement among experts and single-target label noise, in skin lesions, prostate cancer, and retinal disease. Ying et al. \cite{ying2023covid} improved COVID-19 chest X-ray classification through techniques like PCA, low-rank representation, neighborhood graph regularization, and k-nearest neighbor. Similarly, Zhou et al. \cite{zhou2023combating} employed consistency regularization and disentangled distribution learning for multi-label disease classification and severity grading in chest X-rays and diabetic retinopathy. Xue et al. \cite{xue2022robust} combined a student-teacher network with a co-training strategy to improve prostrate cancer grading, skin classification, chest X-ray classification, and histopathology cancer detection, in different label noise settings. Liu et al. \cite{liu2021co} proposed co-correcting, a curriculum learning-based label correction strategy, for robust training with noisy labels in metastatic tissue classification and melanoma classification. 

Despite incorporating some concepts from self-supervised learning, no research has explored the impact of just self-supervised pretraining on enhancing robustness against noisy labels.

\subsection{Self-supervised Pretraining}
\label{self-supervision_pretext_task}
Several self-supervised techniques have emerged recently \cite{gui2023survey}, encompassing simple pretext task-solving approaches \cite{zhang2016colorful,gidaris2018unsupervised,doersch2015unsupervised}, contrastive learning methods \cite{chen2020simple,he2020momentum,zbontar2021barlow}, and generative approaches \cite{liu2021self,zhou2021ibot,he2022masked}. Generative approaches have shown promise but face challenges due to training instability and high computational resource requirements. Additionally, the majority of recent generative approaches necessitate a transformer as the backbone, and investigating the robustness of a transformer-based architecture against label noise, in comparison to a CNN, is a distinct topic of discussion. Furthermore, recent mask image modeling-based generative approaches that learn by randomly masking a certain portion of the image may inadvertently miss crucial features. This issue could be problematic for medical datasets that rely on subtle image cues \cite{huang2023self} and requires a separate investigation. Therefore, for this work, we considered focusing solely on contrastive learning, which is widely used, and the pretext task-based approach, which is simple but unexplored, leaving generative approaches for future investigation.

In this study, we chose three pretext tasks: \textit{Rotation prediction}, \textit{Jigsaw puzzle}, and \textit{Jigmag puzzle}, and a contrastive approach: \textit{SimCLR}. \textit{Rotation prediction} \cite{gidaris2018unsupervised} trains a model to predict the rotation degree of an image in various orientations. \textit{Jigsaw puzzle} \cite{noroozi2016unsupervised} requires training a model to learn to predict the arrangement of shuffled, non-overlapping patches in an image. \textit{Jigmag puzzle} \cite{koohbanani2021self}, originally proposed for histopathology images, learns to predict the arrangement of patches obtained from magnifying an image at various factors. \textit{SimCLR} utilizes a contrastive loss to compare the representations of different augmented views of the same input, aiming to bring closer the augmented views (positive pairs) of the same image while keeping the augmented views of other images (negative pairs) far apart in the representation space. The benefit of pretraining depends on the suitability of the pretraining task with the main task \cite{lee2021predicting}.

We selected these pretext tasks because \textit{Rotation prediction} and \textit{Jigsaw puzzle} are commonly used in the literature, while \textit{Jigmag puzzle} was specifically proposed to address the subtlety of medical images. Other pretext tasks, such as \textit{Colorization} \cite{larsson2017colorization}, were deemed unsuitable for our grayscale X-ray image and stained histopathology image datasets. \textit{SimCLR} was chosen for the contrastive approach due to its simplified framework, which eliminates the need for special memory buffers or specialized architectures.

\section{Datasets}
\subsection{COVID-QU-Ex} This dataset is a collection of chest X-ray images obtained from various patients \cite{anas_2022} categorized into three groups: COVID infection, Non-COVID infection, and Normal. The dataset consists of a total of 27,132 training images, with 8,561 classified as Normal, 9,010 as Non-COVID-19, and 9,561 as COVID-19 cases. Additionally, there is an exclusive test set containing 6,788 images for evaluation.

\subsection{NCT-CRC-HE-100K} This dataset has 100,000 histopathological image patches of size $224 \times 224$ extracted from stained tissue slides \cite{kather2019predicting} featuring nine classes, such as adipose, lymphocytes, mucus, etc. The test set uses a different CRC-VAL-HE-7K dataset consisting of 7,180 images, featuring all nine classes of the training set.
\section{Experimental Setup}

\subsection{Random Label Noise}
To evaluate how a deep learning classifier performs on high label noise, we randomly flipped all the labels in the training set such that the original labels are assigned to any other labels within the close-set with some probability \cite{jiang2018mentornet}. Assuming a training dataset $\{(\mathbf{x}_i, y_i)\}_{i}^{n} \in \mathcal{D}$, which contains $n$ samples, $x_i$ is a data point belonging to the set $\mathcal{X} \in \mathbb{R}^{d}$, and $y_i$ is its corresponding class label from a close-set classes $C = \{c_0,c_1,..,c_4\}$. For any sample $(\mathbf{x}_i, y_i)$, we change its label $y_i$ to $\hat{y_{i}} \overset{p}{\sim} C \setminus y_{i}$, where $p$ is the noise probability and $C \setminus y_{i}$ denotes any label of close-set classes other than the true label. The label noise is symmetrical for all the classes within the close set. We conducted experiments using four different noise rates $p \in \{0.5,0.6,0.7,0.8\}$. 

\begin{figure}[h!]
\centering
\begin{subfigure}[t]{0.48\textwidth}
    \includegraphics[width=1\linewidth]{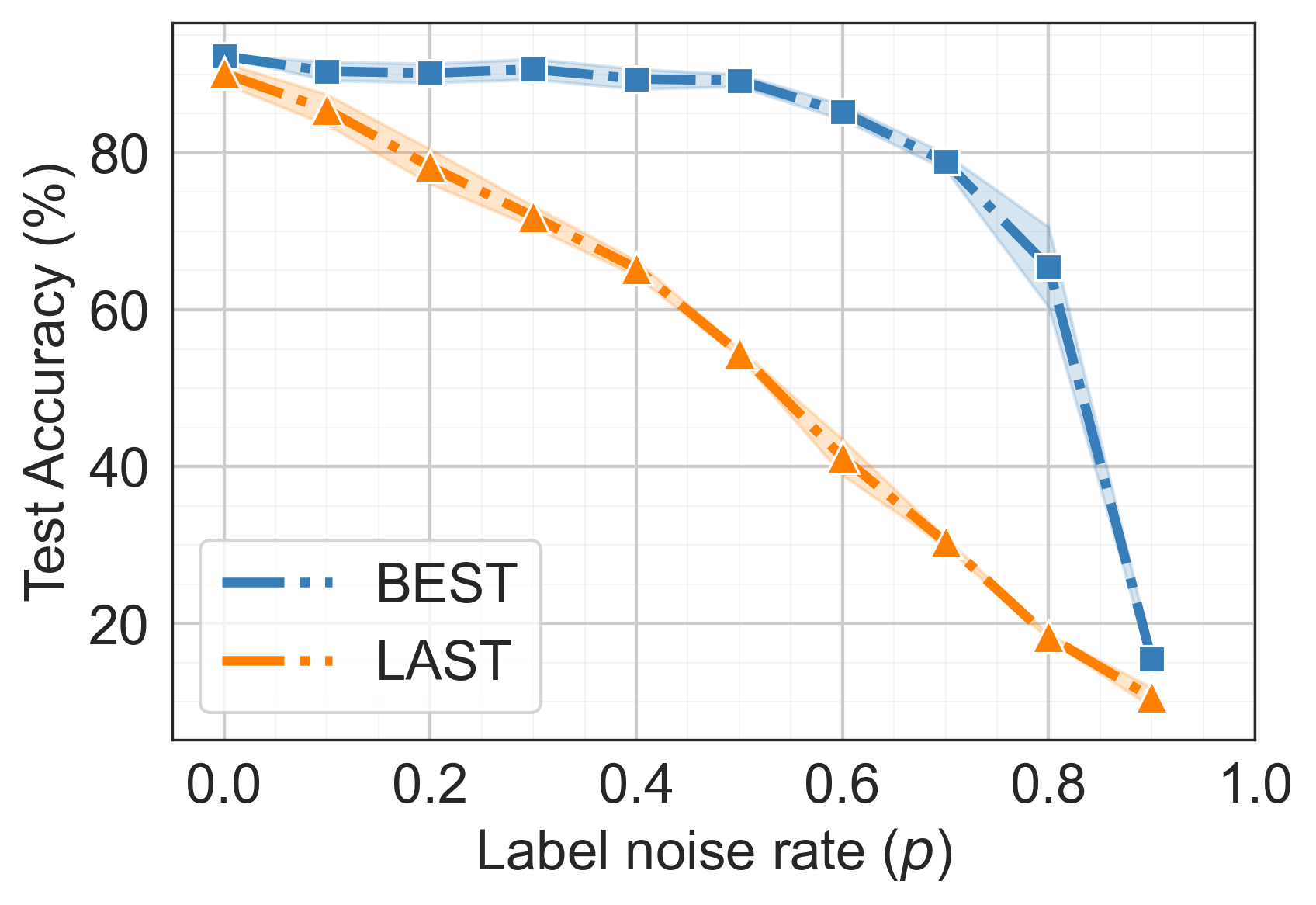}
    \label{fig:histopath_noise}
    \caption{NCT-CRC-HE-100K}
  \end{subfigure}
\begin{subfigure}[t]{0.48\textwidth}
    \includegraphics[width=1\linewidth]{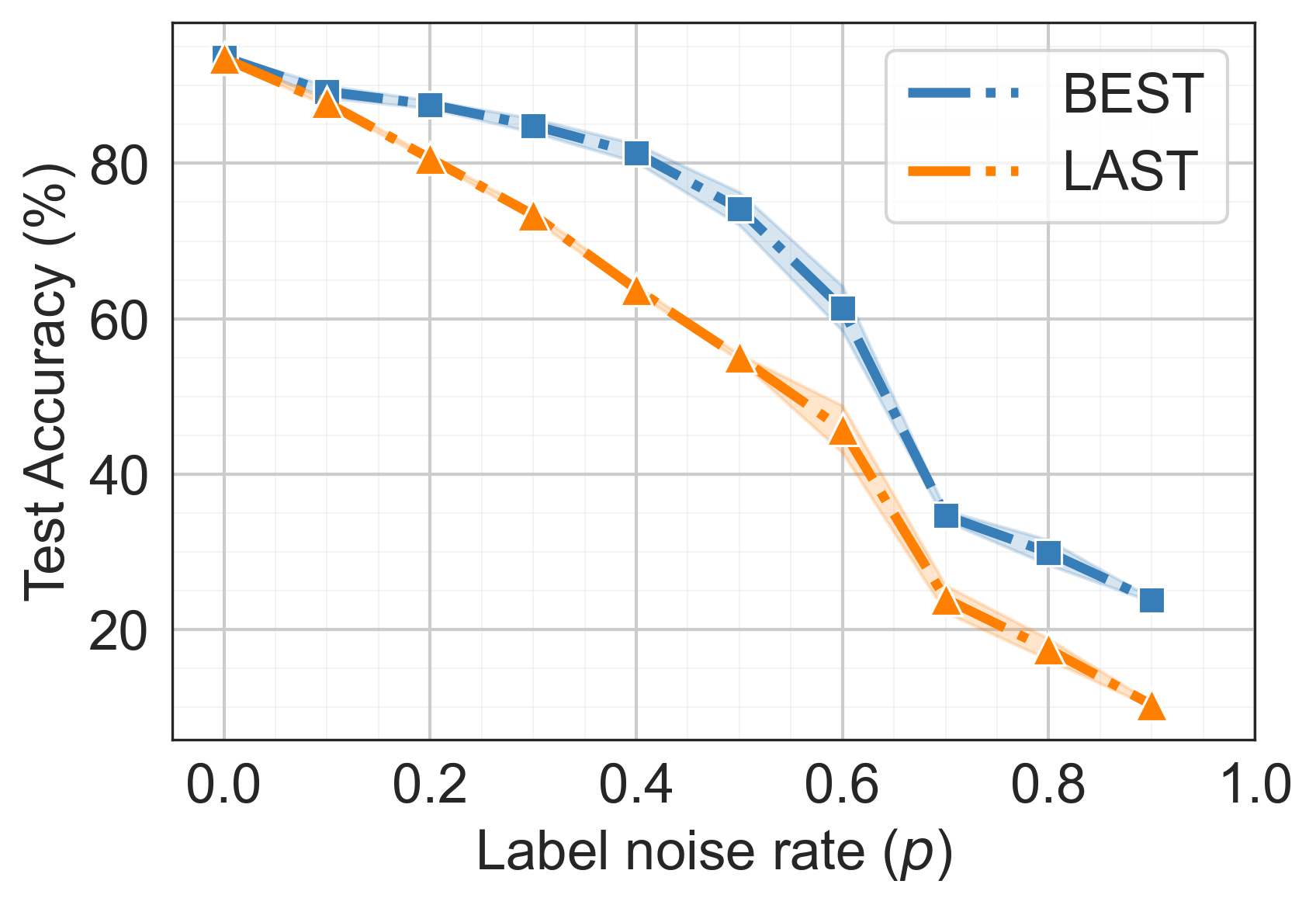}
    \label{fig:covid19_noise}
    \caption{COVID-QU-Ex}
  \end{subfigure}
\caption{Test performance as a function of training label noise rate (noise probability \textit{p}) ranging from scale 0 to 1, with the shaded region indicating variability across three experimental trials. BEST indicates the highest test accuracy achieved, while LAST denotes the average test accuracy achieved in the last five epochs.}
\label{fig:noise_impact}
\end{figure}

As depicted in Fig. \ref{fig:noise_impact}, the impact of noisy labels on test performance varies across datasets; \textit{NCT-CRC-HE-100K} remains robust to noise below 0.5, whereas \textit{COVID-QU-Ex} is affected at lower rates also.

\subsection{Methodology}

Our approach involves two stages: i) pretrain a model using self-supervised learning on the given dataset to learn meaningful feature extractors, and ii) train the pretrained model for medical image classification on the same dataset with noisy labels (Fig. \ref{fig:pipeline}). We primarily focus on the first stage, experimenting with four self-supervised tasks. In the second stage, we experimented with cross-entropy alone, followed by two state-of-the-art LNL approaches, Co-teaching \cite{han2018co} and DivideMix \cite{Li2020DivideMix:}. 

Co-teaching selectively samples clean examples by ranking the training loss, while DivideMix utilizes a Gaussian Mixture Model (GMM) to categorize examples into clean and noisy groups based on the training loss of each sample. DivideMix also applies the MixMatch \cite{berthelot2019mixmatch} semi-supervised learning approach by treating noisy labels as unlabeled examples. Notably, Co-teaching focuses on clean sample selection, while DivideMix does both clean sample selection and noisy label correction, but both use dual networks and utilize a warm-up phase.
 
\paragraph{Evaluation:}
Following \cite{han2018co,Li2020DivideMix:}, we evaluate the best test classification accuracy (BEST) and the average test accuracy of the last five epochs (LAST). The test set serves as a pseudo-test set, accessing the model's maximum performance with BEST, while LAST measures if the model has overfitted to noisy labels (see Fig. \ref{fig:noise_impact}).
\subsection{Implementation Details}

\subsubsection{Self-supervised Pretraining:}
\label{selfsupervised_implementation}
We utilized the ResNet18 architecture for all experiments. For Rotation prediction, images were resized and underwent strong data augmentations: random horizontal flips, small rotations ($10^{\circ}$), sharpness adjustment, equalization, and auto contrast. The model had to predict the rotation angle from four possible angles ($0^{\circ},90^{\circ},180^{\circ}, 270^{\circ}$). 

For Jigsaw puzzle solving, we performed similar strong augmentations and divided the resized input image into a $3\times3$ grid of patches. The patches were resized to $64 \times 64$ pixels, normalized with patch mean and standard deviation, and randomly shuffled to create one of the 1000 chosen permutations \footnote{https://github.com/bbrattoli/JigsawPuzzlePytorch}. Then, they were passed through the ResNet18 feature extractor and concatenated before being fed into a fully connected output layer that predicted the permutation. 

For Jigmag puzzle solving, we applied the aforementioned augmentations and randomly magnified the input image at different locations using nine magnification factors ranging from 1 to 5. The magnified patches were resized, normalized, and randomly rearranged into one of the 1000 chosen permutations similar to the Jigsaw puzzle. A fully connected softmax layer after the ResNet feature extractor predicted the permutation.
We used SimCLR \cite{chen2020simple} implemented in \footnote{https://github.com/sthalles/SimCLR}, setting the default parameters. The input images were resized and augmented with random horizontal flips, color jitter, and Gaussian blur. Table \ref{self-supervised hyperparameter} summarizes all training hyperparameter settings. All the methods were trained until convergence, with the number of epochs and learning rates chosen accordingly. For instance, Rotation prediction performed best with an SGD learning rate of 0.01 compared to other settings, while Jigsaw and Jigmag converged effectively using a batch size of 128. \begin{table}[]
\scriptsize
\centering
\caption{Hyperparameters used for training various self-supervised methods.}
\label{self-supervised hyperparameter}
\begin{tabular}{l|l|l|l|l|l|l|l|l}
\hline
Datasets                         & Method & Input size    & Batch & Epochs & Wt decay & Lr & Optim & Sheduler         \\ \hline
\multirow{4}{*}{NCT-CRC-HE-100K} & Rotation                                                        & $224 \times 224$     & $256$        & $70$     & $10^{-4}$         & $0.01$                                                             & SGD       & Cosine Annealing\\ \cline{2-9} 
                                 & Jigsaw                                                          & $64 \times 64$ & $128$        & $50$     & $10^{-4}$         & $0.001$                                                            & Adam      & Cosine Annealing \\ \cline{2-9} 
                                 & Jigmag                                                          & $64 \times 64$ & $128$        & $50$     & $10^{-4}$         & $0.001$                                                            & Adam      & Cosine Annealing \\ \cline{2-9} 
                                 & SimCLR                                                          & $224 \times 224$     & $256$        & $100$    & $10^{-4}$         & $0.001$                                                            & Adam      & Cosine Annealing \\ \hline
\multirow{4}{*}{COVID-QU-Ex}     & Rotation                                                        & $224 \times 224$     & $256$        & $70$     & $10^{-4}$         & $0.01$                                                             & SGD       & Cosine Annealing \\ \cline{2-9} 
                                 & Jigsaw                                                          & $64 \times 64$ & $128$        & $60$     & $10^{-4}$         & $0.001$                                                            & Adam      & Cosine Annealing \\ \cline{2-9} 
                                 & Jigmag                                                          & $64 \times 64$ & $128$        & $60$     & $10^{-4}$         & $0.001$                                                            & Adam      & Cosine Annealing \\ \cline{2-9} 
                                 & SimCLR                                                          & $224 \times 224$     & $256$        & $200$    & $10^{-4}$         & $0.001$                                                            & Adam      & Cosine Annealing \\ \hline
\end{tabular}
\vspace{-2em}
\end{table}

\subsubsection{Learning with Noisy Labels:}
In this stage, we initialized the ResNet18 feature extractor with weights from self-supervised training, added a fully connected output layer, and retrained the entire model with noisy labels. For the first experiment, we trained the model using standard cross-entropy loss without any modifications. The training process involved a batch size of 256, an SGD optimizer with a momentum of 0.9, weight decay of $10^{-4}$, an initial learning rate of 0.01, and 50 training epochs.

For Co-teaching, we followed the original paper's \cite{han2018co} recommendations and set the warm-up epochs to 10, $\tau = p$ and $c = 1$, where $p$ is the label noise rate in data. As for DivideMix, we slightly adjusted the original hyperparameters \cite{Li2020DivideMix:}, setting the warm-up epochs to 10, $M = 2$, $T = 0.2$, $\alpha = 4$, $\tau = 0.2$, and $\lambda_{u} = 0$ for $p = \{0.5,0.6,0.7\}$, while $\lambda_{u}$ was changed to $0.25$ for $p=0.8$. Both methods maintained other training hyperparameters the same as the standard cross-entropy approach, except for DivideMix, where a batch size of 128 was used. To avoid confirmation bias, both Co-teaching and DivideMix original implementations initialize the dual networks with different weights. Similarly, in our approach, we adopt this strategy by initializing the dual networks with two distinct pretrained weights obtained from separate self-supervised training under the same settings. 

All our experiments were implemented in Python 3.8 using the PyTorch 12.1.1 framework and trained on an A100 GPU (40 GB). We ran 3 experimental trials for each case to report the mean and standard deviation. 

\section{Results}

\subsubsection{Self-supervised Pretraining improves Robustness against Noisy Labels:}
\begin{figure}[h!]
\centering
\includegraphics[width=1\linewidth]{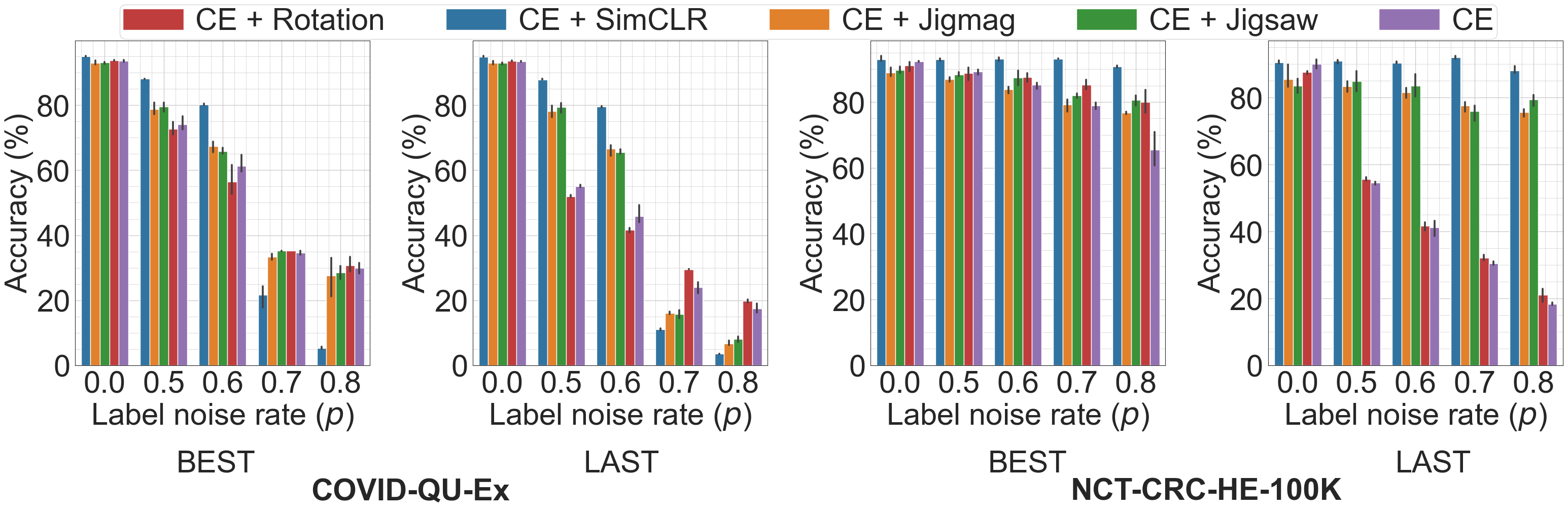}
\caption{Performance comparison of models trained starting from default weights vs. trained from weights initialized from self-supervised pretraining, when using standard cross entropy (CE) loss at different training label noise rates ($p$), in \textit{COVID-QU-Ex} and \textit{NCT-CRC-HE-100K}. BEST denotes the best test accuracy, while LAST denotes the average test accuracy achieved of the last five epochs. The experiments were run for three experimental trials to report the error bar. }
\label{fig:finetune_cross_entropy}
\end{figure}

In Fig. \ref{fig:finetune_cross_entropy}, we compared models trained with standard cross-entropy (CE) loss, using weights initialized from the pretext task against PyTorch's default randomized He initialization \cite{he2015delving}. 

The results demonstrate that self-supervised pretrained models significantly improve in terms of the BEST and LAST accuracy, particularly at high noise rates in the \textit{NCT-CRC-HE-100K}. Specifically, SimCLR achieved better performance in both BEST and LAST accuracy at all noise rates, while Jigsaw and Jigmag also notably improve the LAST accuracy at $p = \{0.6,0.7,0.8\}$.
 
Similar trends are observed in the \textit{COVID-QU-Ex}, where SimCLR, Jigmag, and Jigsaw outperform others significantly at $p = \{0.5,0.6\}$ in terms of both BEST and LAST accuracy. At $p = \{0.7,0.8\}$, rotation performs better, but the improvements are not as good as those observed with SimCLR, Jigsaw, and Jigmag in the $p = \{0.5,0.6\}$ range. However, SimCLR, Jigmag, and Jigsaw performed worst than the cross entropy in the range $p = \{0.7,0.8\}$. 

\textit{The choice of the best self-supervised task varies based on the dataset, noise rate, and evaluation criteria, but the results achieved using self-supervised pretraining consistently show better performance compared to directly starting training from PyTorch's default randomized He initialization}.

\subsubsection{Self-supervised Pretraining for LNL methods:}
\begin{figure}[h!]
\centering
\begin{subfigure}[t]{1\textwidth}
    \includegraphics[width=1\linewidth]{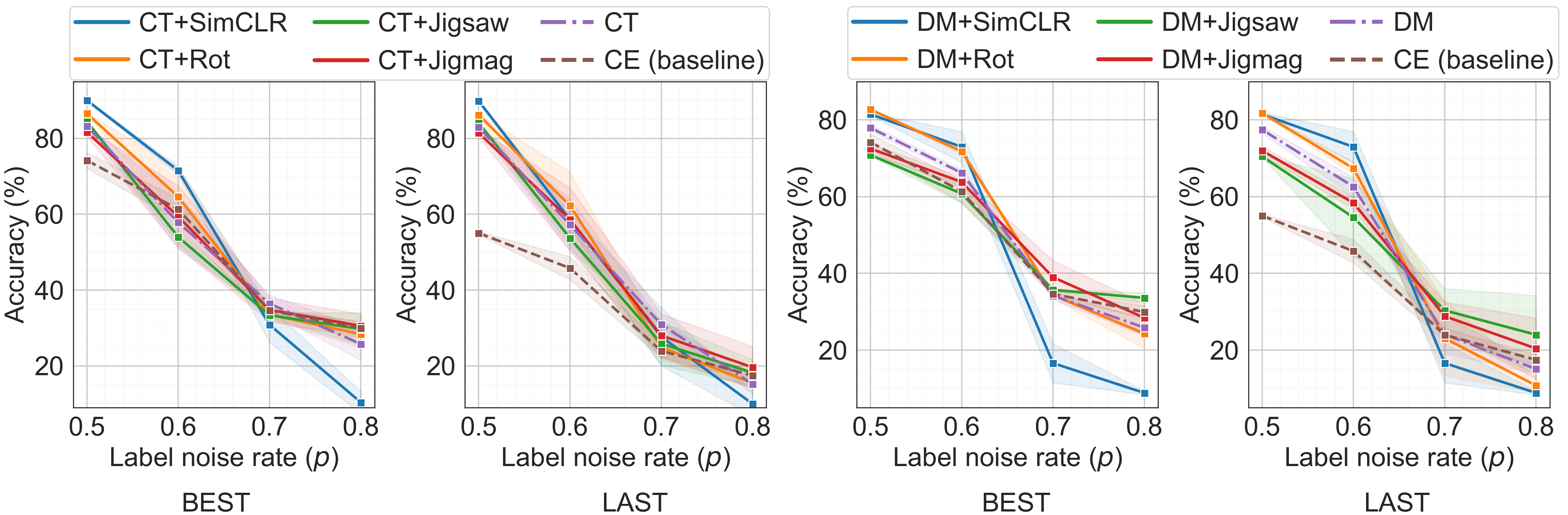}
    \caption{COVID-QU-Ex}
    \vspace{1em}
  \end{subfigure}
\begin{subfigure}[t]{1\textwidth}
    \includegraphics[width=1\linewidth]{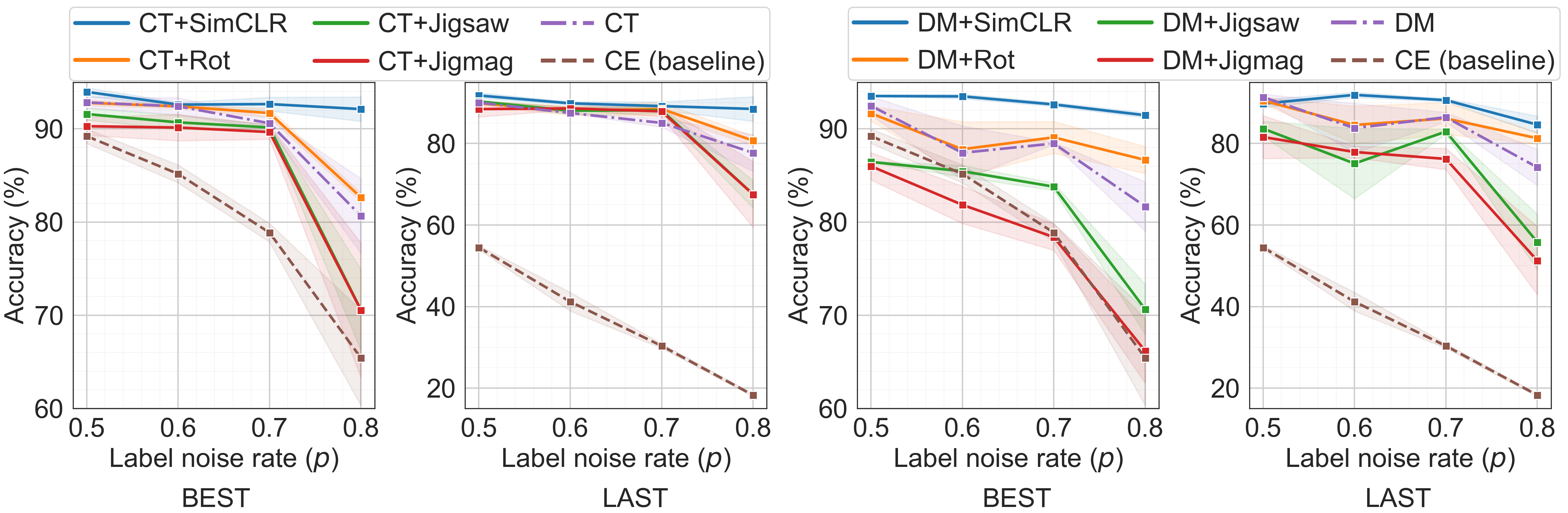}
    \caption{NCT-CRC-HE-100K}
  \end{subfigure}
\caption{Performance comparison of existing LNL methods when initialized with self-supervised pretraining against baselines at different training label noise rates ($p$), in (a) \textit{COVID-QU-Ex} and (b) \textit{NCT-CRC-HE-100K}. Cross entropy (CE) denotes the actual baseline, Coteaching (CT) and Dividemix (DM) are existing LNL methods, and the term after $+$ represents various self-supervised pretraining methods. BEST denotes the best test accuracy, while LAST denotes the average test accuracy achieved of the last five epochs. The shaded region along the line indicates the variability across three experimental trials.}
\label{fig:histopathology_and_covid_all}
\end{figure}

We compared LNL methods trained from PyTorch's default He initialization with those trained using weights initialized from a self-supervised pretraining in Fig.\ref{fig:histopathology_and_covid_all}. The results show that the LNL methods already achieved good performance in the \textit{NCT-CRC-HE-100K}, with only a small room for improvement at a lower noise rate. At higher noise rates, SimCLR achieved the highest BEST and LAST accuracy, followed by Rotation prediction. Initializing LNL with Jigsaw and Jigmag didn't improve the performance, but rather degraded it.

In the \textit{COVID-QU-Ex}, we observed a significant performance boost in the noise range $p = \{0.5, 0.6\}$ for both Coteaching and DivideMix when using pretrained weights from the SimCLR and Rotation prediction task. However, beyond $p = \{0.7, 0.8\}$, the scores exhibited high variability, making it difficult to identify the best performance. SimCLR struggled with high noise rates, possibly due to excessive contrastive learning augmentations that overlooked vital subtle features needed for maintaining performance in X-ray images with high label noise. Additionally, in \textit{COVID-QU-Ex}, we noticed that the clean samples selected by LNL methods were biased towards one class and ignored the other classes, particularly at high noise levels, making it intriguing to investigate the cause behind this phenomenon in the future.

\section{Conclusion}
We examined the effectiveness of utilizing self-supervised pretraining alone to improve the model's robustness against noisy labels in medical image classification. The choice of self-supervised task varied depending on the dataset, noise rate, and, evaluation criteria, with SimCLR consistently yielding the best results in most cases with LNL. 

This study addresses a gap in the current research by serving as a first demonstration of the benefits of various self-supervised pretraining for medical image classification with noisy labels and offering valuable insights to mitigate the impact of high label noise. In the future, we plan to investigate additional self-supervised baselines and further explore how the nature and size of the dataset influence the improvements offered by self-supervised pretraining in terms of robustness against noisy labels.

Additionally, in this work, we have limited the investigation to CNN-based architecture. It would be interesting to investigate how recent transformer-based architectures behave under various levels of label noise, whether off-the-shelf LNL methods work with transformer architecture, and how transformer-based self-supervised techniques improve robustness against noisy labels.
\\\\
\noindent\textbf{Acknowledgements.} Research reported in this publication was supported by the National Institute of General Medical Sciences Award No. R35GM128877 of the National Institutes of Health, the Office of Advanced Cyber Infrastructure Award No. 1808530 of the National Science Foundation, and the Division Of Chemistry, Bioengineering, Environmental, and Transport Systems Award No. 2245152 of the National Science Foundation. We would like to thank the Research Computing team \cite{RITRC} at the Rochester Institute of Technology for proving computing resources for this research.

\bibliographystyle{plain}
\bibliography{mybib}

\begin{thebibliography}{10}

\bibitem{berthelot2019mixmatch}
David Berthelot, Nicholas Carlini, Ian Goodfellow, Nicolas Papernot, Avital
  Oliver, and Colin~A Raffel.
\newblock Mixmatch: A holistic approach to semi-supervised learning.
\newblock {\em Advances in neural information processing systems}, 32, 2019.

\bibitem{chen2019understanding}
Pengfei Chen, Ben~Ben Liao, Guangyong Chen, and Shengyu Zhang.
\newblock Understanding and utilizing deep neural networks trained with noisy
  labels.
\newblock In {\em International Conference on Machine Learning}, pages
  1062--1070. PMLR, 2019.

\bibitem{chen2020simple}
Ting Chen, Simon Kornblith, Mohammad Norouzi, and Geoffrey Hinton.
\newblock A simple framework for contrastive learning of visual
  representations.
\newblock In {\em International conference on machine learning}, pages
  1597--1607. PMLR, 2020.

\bibitem{dgani2018training}
Yair Dgani, Hayit Greenspan, and Jacob Goldberger.
\newblock Training a neural network based on unreliable human annotation of
  medical images.
\newblock In {\em 2018 IEEE 15th International symposium on biomedical imaging
  (ISBI 2018)}, pages 39--42. IEEE, 2018.

\bibitem{doersch2015unsupervised}
Carl Doersch, Abhinav Gupta, and Alexei~A Efros.
\newblock Unsupervised visual representation learning by context prediction.
\newblock In {\em Proceedings of the IEEE international conference on computer
  vision}, pages 1422--1430, 2015.

\bibitem{gidaris2018unsupervised}
Spyros Gidaris, Praveer Singh, and Nikos Komodakis.
\newblock Unsupervised representation learning by predicting image rotations.
\newblock {\em arXiv preprint arXiv:1803.07728}, 2018.

\bibitem{gui2023survey}
Jie Gui, Tuo Chen, Qiong Cao, Zhenan Sun, Hao Luo, and Dacheng Tao.
\newblock A survey of self-supervised learning from multiple perspectives:
  Algorithms, theory, applications and future trends.
\newblock {\em arXiv preprint arXiv:2301.05712}, 2023.

\bibitem{han2018co}
Bo~Han, Quanming Yao, Xingrui Yu, Gang Niu, Miao Xu, Weihua Hu, Ivor Tsang, and
  Masashi Sugiyama.
\newblock Co-teaching: Robust training of deep neural networks with extremely
  noisy labels.
\newblock {\em Advances in neural information processing systems}, 31, 2018.

\bibitem{he2022masked}
Kaiming He, Xinlei Chen, Saining Xie, Yanghao Li, Piotr Doll{\'a}r, and Ross
  Girshick.
\newblock Masked autoencoders are scalable vision learners.
\newblock In {\em Proceedings of the IEEE/CVF Conference on Computer Vision and
  Pattern Recognition}, pages 16000--16009, 2022.

\bibitem{he2020momentum}
Kaiming He, Haoqi Fan, Yuxin Wu, Saining Xie, and Ross Girshick.
\newblock Momentum contrast for unsupervised visual representation learning.
\newblock In {\em Proceedings of the IEEE/CVF conference on computer vision and
  pattern recognition}, pages 9729--9738, 2020.

\bibitem{he2015delving}
Kaiming He, Xiangyu Zhang, Shaoqing Ren, and Jian Sun.
\newblock Delving deep into rectifiers: Surpassing human-level performance on
  imagenet classification.
\newblock In {\em Proceedings of the IEEE international conference on computer
  vision}, pages 1026--1034, 2015.

\bibitem{hu2019simple}
Wei Hu, Zhiyuan Li, and Dingli Yu.
\newblock Simple and effective regularization methods for training on noisily
  labeled data with generalization guarantee.
\newblock {\em arXiv preprint arXiv:1905.11368}, 2019.

\bibitem{huang2023self}
Shih-Cheng Huang, Anuj Pareek, Malte Jensen, Matthew~P Lungren, Serena Yeung,
  and Akshay~S Chaudhari.
\newblock Self-supervised learning for medical image classification: a
  systematic review and implementation guidelines.
\newblock {\em NPJ Digital Medicine}, 6(1):74, 2023.

\bibitem{jiang2018mentornet}
Lu~Jiang, Zhengyuan Zhou, Thomas Leung, Li-Jia Li, and Li~Fei-Fei.
\newblock Mentornet: Learning data-driven curriculum for very deep neural
  networks on corrupted labels.
\newblock In {\em International conference on machine learning}, pages
  2304--2313. PMLR, 2018.

\bibitem{ju2022improving}
Lie Ju, Xin Wang, Lin Wang, Dwarikanath Mahapatra, Xin Zhao, Quan Zhou,
  Tongliang Liu, and Zongyuan Ge.
\newblock Improving medical images classification with label noise using
  dual-uncertainty estimation.
\newblock {\em IEEE transactions on medical imaging}, 41(6):1533--1546, 2022.

\bibitem{karimi2020deep}
Davood Karimi, Haoran Dou, Simon~K Warfield, and Ali Gholipour.
\newblock Deep learning with noisy labels: Exploring techniques and remedies in
  medical image analysis.
\newblock {\em Medical image analysis}, 65:101759, 2020.

\bibitem{kather2019predicting}
Jakob~Nikolas Kather, Johannes Krisam, Pornpimol Charoentong, Tom Luedde,
  Esther Herpel, Cleo-Aron Weis, Timo Gaiser, Alexander Marx, Nektarios~A
  Valous, Dyke Ferber, et~al.
\newblock Predicting survival from colorectal cancer histology slides using
  deep learning: A retrospective multicenter study.
\newblock {\em PLoS Medicine}, 16(1):e1002730, 2019.

\bibitem{khanal2023investigating}
Bidur Khanal, SM~Kamrul Hasan, Bishesh Khanal, and Cristian~A Linte.
\newblock Investigating the impact of class-dependent label noise in medical
  image classification.
\newblock In {\em Medical Imaging 2023: Image Processing}, volume 12464, pages
  728--733. SPIE, 2023.

\bibitem{khanal2021does}
Bidur Khanal and Christopher Kanan.
\newblock How does heterogeneous label noise impact generalization in neural
  nets?
\newblock In {\em Advances in Visual Computing: 16th International Symposium,
  ISVC 2021, Virtual Event, October 4-6, 2021, Proceedings, Part II}, pages
  229--241. Springer, 2021.

\bibitem{koohbanani2021self}
Navid~Alemi Koohbanani, Balagopal Unnikrishnan, Syed~Ali Khurram, Pavitra
  Krishnaswamy, and Nasir Rajpoot.
\newblock Self-path: Self-supervision for classification of pathology images
  with limited annotations.
\newblock {\em IEEE Transactions on Medical Imaging}, 40(10):2845--2856, 2021.

\bibitem{larsson2017colorization}
Gustav Larsson, Michael Maire, and Gregory Shakhnarovich.
\newblock Colorization as a proxy task for visual understanding.
\newblock In {\em Proceedings of the IEEE conference on computer vision and
  pattern recognition}, pages 6874--6883, 2017.

\bibitem{le2019pancreatic}
Han Le, Dimitris Samaras, Tahsin Kurc, Rajarsi Gupta, Kenneth Shroyer, and Joel
  Saltz.
\newblock Pancreatic cancer detection in whole slide images using noisy label
  annotations.
\newblock In {\em Medical Image Computing and Computer Assisted
  Intervention--MICCAI 2019: 22nd International Conference, Shenzhen, China,
  October 13--17, 2019, Proceedings, Part I 22}, pages 541--549. Springer,
  2019.

\bibitem{lee2021predicting}
Jason~D Lee, Qi~Lei, Nikunj Saunshi, and Jiacheng Zhuo.
\newblock Predicting what you already know helps: Provable self-supervised
  learning.
\newblock {\em Advances in Neural Information Processing Systems}, 34:309--323,
  2021.

\bibitem{lee2019robust}
Kimin Lee, Sukmin Yun, Kibok Lee, Honglak Lee, Bo~Li, and Jinwoo Shin.
\newblock Robust inference via generative classifiers for handling noisy
  labels.
\newblock In {\em International conference on machine learning}, pages
  3763--3772. PMLR, 2019.

\bibitem{Li2020DivideMix:}
Junnan Li, Richard Socher, and Steven~C.H. Hoi.
\newblock Dividemix: Learning with noisy labels as semi-supervised learning.
\newblock In {\em International Conference on Learning Representations}, 2020.

\bibitem{liu2021co}
Jiarun Liu, Ruirui Li, and Chuan Sun.
\newblock Co-correcting: noise-tolerant medical image classification via mutual
  label correction.
\newblock {\em IEEE Transactions on Medical Imaging}, 40(12):3580--3592, 2021.

\bibitem{liu2020early}
Sheng Liu, Jonathan Niles-Weed, Narges Razavian, and Carlos Fernandez-Granda.
\newblock Early-learning regularization prevents memorization of noisy labels.
\newblock {\em Advances in neural information processing systems},
  33:20331--20342, 2020.

\bibitem{liu2021self}
Xiao Liu, Fanjin Zhang, Zhenyu Hou, Li~Mian, Zhaoyu Wang, Jing Zhang, and Jie
  Tang.
\newblock Self-supervised learning: Generative or contrastive.
\newblock {\em IEEE transactions on knowledge and data engineering},
  35(1):857--876, 2021.

\bibitem{noroozi2016unsupervised}
Mehdi Noroozi and Paolo Favaro.
\newblock Unsupervised learning of visual representations by solving jigsaw
  puzzles.
\newblock In {\em Computer Vision--ECCV 2016: 14th European Conference,
  Amsterdam, The Netherlands, October 11-14, 2016, Proceedings, Part VI}, pages
  69--84. Springer, 2016.

\bibitem{pham2021interpreting}
Hieu~H Pham, Tung~T Le, Dat~Q Tran, Dat~T Ngo, and Ha~Q Nguyen.
\newblock Interpreting chest x-rays via cnns that exploit hierarchical disease
  dependencies and uncertainty labels.
\newblock {\em Neurocomputing}, 437:186--194, 2021.

\bibitem{RITRC}
{Rochester Institute of Technology}.
\newblock Research computing services, 2022.

\bibitem{song2019does}
Hwanjun Song, Minseok Kim, Dongmin Park, and Jae-Gil Lee.
\newblock How does early stopping help generalization against label noise?
\newblock {\em arXiv preprint arXiv:1911.08059}, 2019.

\bibitem{anas_2022}
Anas~M. Tahir et~al.
\newblock {COVID-QU-Ex Dataset}, 2022.

\bibitem{Wei_2020_CVPR}
Hongxin Wei, Lei Feng, Xiangyu Chen, and Bo~An.
\newblock Combating noisy labels by agreement: A joint training method with
  co-regularization.
\newblock In {\em Proceedings of the IEEE/CVF Conference on Computer Vision and
  Pattern Recognition (CVPR)}, June 2020.

\bibitem{xue2019robust}
Cheng Xue, Qi~Dou, Xueying Shi, Hao Chen, and Pheng-Ann Heng.
\newblock Robust learning at noisy labeled medical images: Applied to skin
  lesion classification.
\newblock In {\em 2019 IEEE 16th International symposium on biomedical imaging
  (ISBI 2019)}, pages 1280--1283. IEEE, 2019.

\bibitem{xue2022robust}
Cheng Xue, Lequan Yu, Pengfei Chen, Qi~Dou, and Pheng-Ann Heng.
\newblock Robust medical image classification from noisy labeled data with
  global and local representation guided co-training.
\newblock {\em IEEE Transactions on Medical Imaging}, 41(6):1371--1382, 2022.

\bibitem{ying2023covid}
Xiaoqing Ying, Hao Liu, and Rong Huang.
\newblock Covid-19 chest x-ray image classification in the presence of noisy
  labels.
\newblock {\em Displays}, page 102370, 2023.

\bibitem{zbontar2021barlow}
Jure Zbontar, Li~Jing, Ishan Misra, Yann LeCun, and St{\'e}phane Deny.
\newblock Barlow twins: Self-supervised learning via redundancy reduction.
\newblock In {\em International Conference on Machine Learning}, pages
  12310--12320. PMLR, 2021.

\bibitem{zhang2021understanding}
Chiyuan Zhang, Samy Bengio, Moritz Hardt, Benjamin Recht, and Oriol Vinyals.
\newblock Understanding deep learning (still) requires rethinking
  generalization.
\newblock {\em Communications of the ACM}, 64(3):107--115, 2021.

\bibitem{zhang2016colorful}
Richard Zhang, Phillip Isola, and Alexei~A Efros.
\newblock Colorful image colorization.
\newblock In {\em Computer Vision--ECCV 2016: 14th European Conference,
  Amsterdam, The Netherlands, October 11-14, 2016, Proceedings, Part III 14},
  pages 649--666. Springer, 2016.

\bibitem{zhang2021codim}
Xin Zhang, Zixuan Liu, Kaiwen Xiao, Tian Shen, Junzhou Huang, Wei Yang,
  Dimitris Samaras, and Xiao Han.
\newblock Codim: Learning with noisy labels via contrastive semi-supervised
  learning.
\newblock {\em arXiv preprint arXiv:2111.11652}, 2021.

\bibitem{zheltonozhskii2022contrast}
Evgenii Zheltonozhskii, Chaim Baskin, Avi Mendelson, Alex~M Bronstein, and
  Or~Litany.
\newblock Contrast to divide: Self-supervised pre-training for learning with
  noisy labels.
\newblock In {\em Proceedings of the IEEE/CVF Winter Conference on Applications
  of Computer Vision}, pages 1657--1667, 2022.

\bibitem{zhou2021ibot}
Jinghao Zhou, Chen Wei, Huiyu Wang, Wei Shen, Cihang Xie, Alan Yuille, and Tao
  Kong.
\newblock ibot: Image bert pre-training with online tokenizer.
\newblock {\em arXiv preprint arXiv:2111.07832}, 2021.

\bibitem{zhou2023combating}
Yi~Zhou, Lei Huang, Tao Zhou, and Hanshi Sun.
\newblock Combating medical noisy labels by disentangled distribution learning
  and consistency regularization.
\newblock {\em Future Generation Computer Systems}, 141:567--576, 2023.

\end{thebibliography}
\chapter*{Supplementary Materials}

\textbf{Qualitative Analysis of learned CNN Filters:}\\

\noindent The visualization of filters in the first layer allows us to qualitatively analyze the meaningfulness and alignment of early feature layers learned during the self-supervised task with those of the downstream classification task. When the early filters learned by the task closely match those of the downstream task, we can expect better transferability to downstream tasks. Additionally, initializing the model with such filters before training on noisy labels would reduce the risk of the model learning corrupted features.

\begin{figure}[h!]
\centering
\includegraphics[width=1\linewidth]{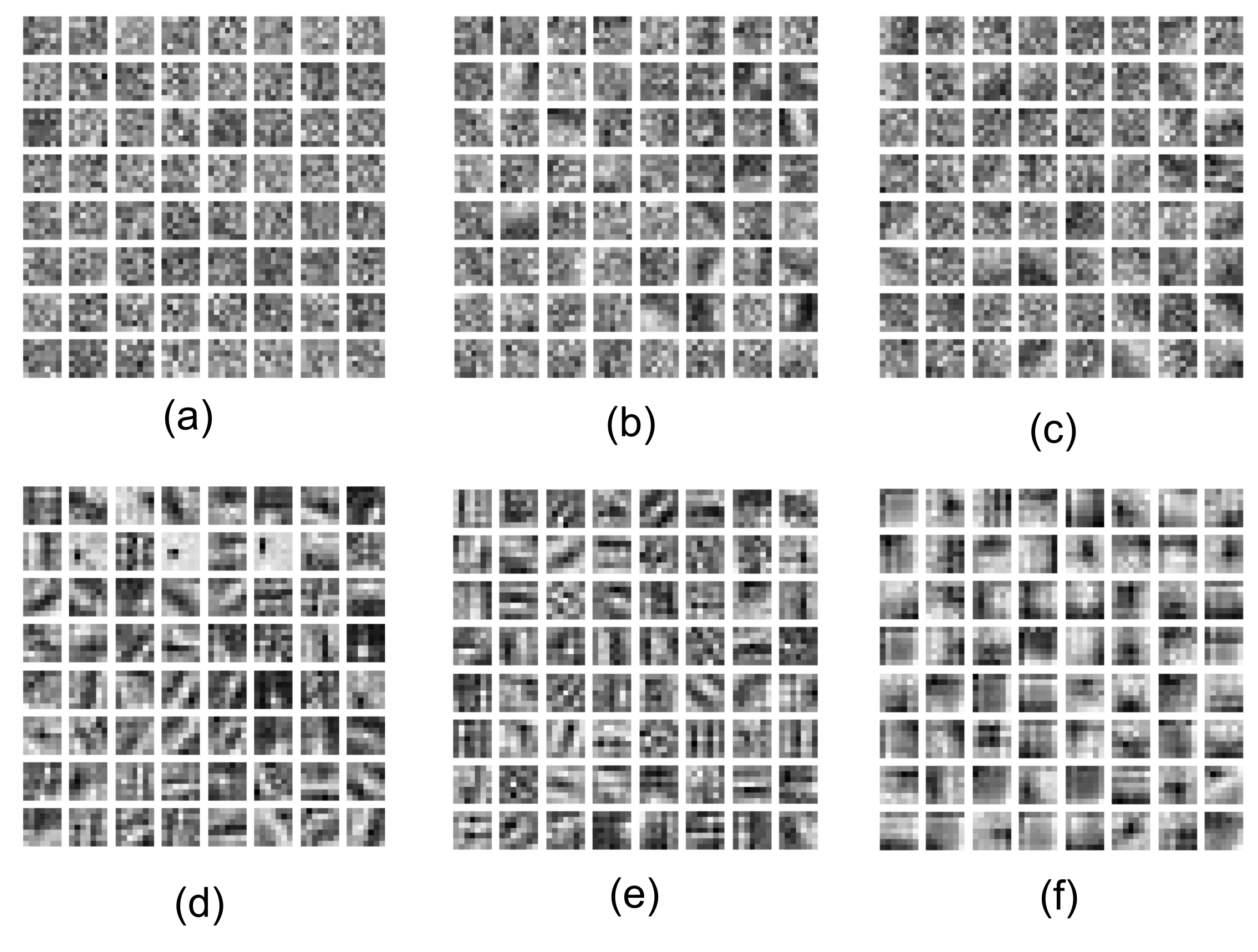}
\vspace{-1em}
\caption{Comparision of 16 randomly chosen first layer filters of ResNet18 after training on COVID-QU-Ex Dataset. (a) He initialized filters, (b) filters learned after supervised training using all the clean labels, (c) filters learned by the Rotation prediction, d) filters learned by solving the Jigsaw puzzle, (e) filters learned by solving the Jigmag puzzle, and (f) filters learned by SimCLR.}
\label{fig:covid_dataset_filters}
\end{figure}

\begin{figure}[h!]
\centering
\includegraphics[width=1\linewidth]{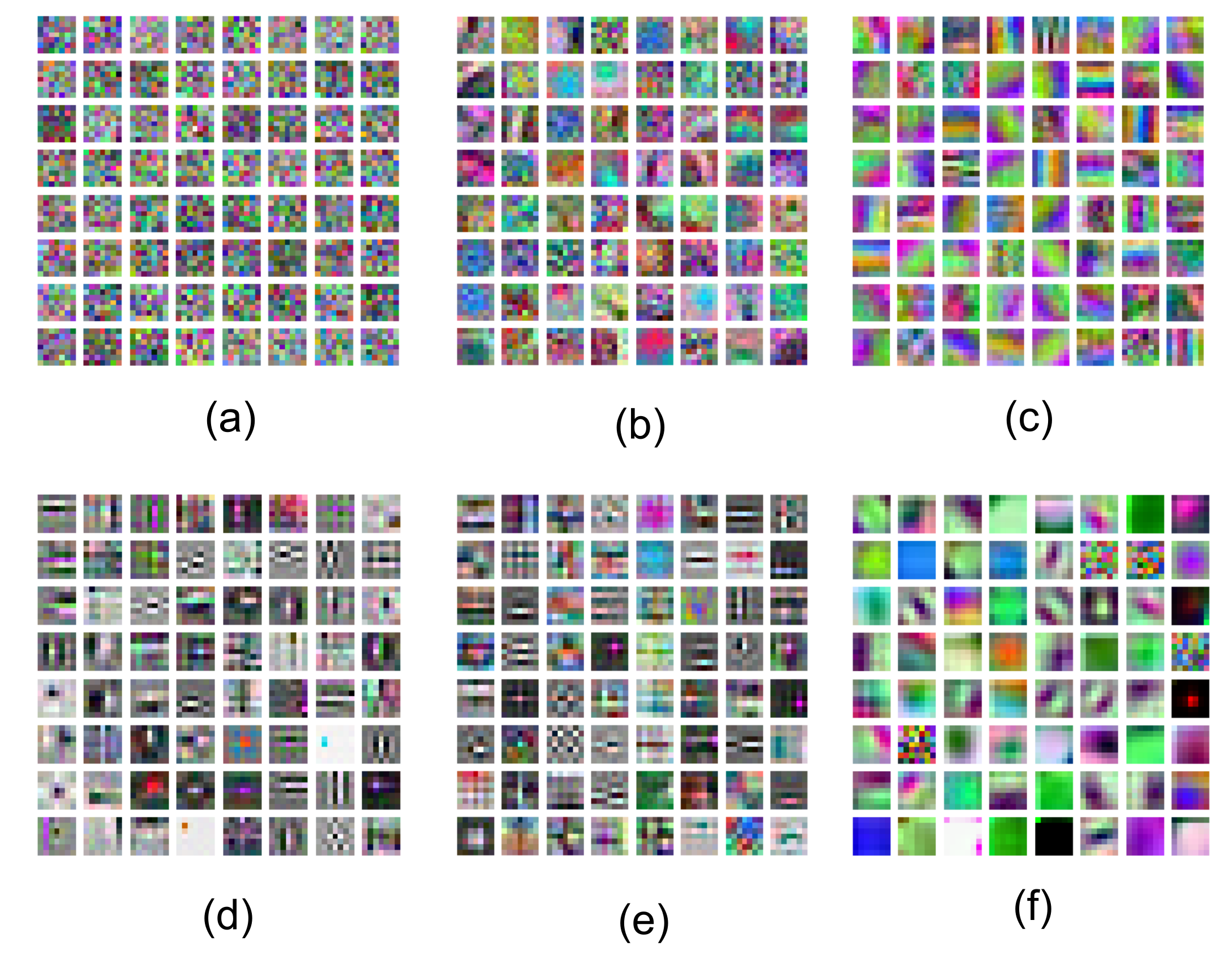}
\vspace{-1em}
\caption{Comparison of 16 randomly chosen first layer filters of ResNet18 after training on NCT-CRC-HE-100K Dataset. (a) He initialized filters, (b) filters learned after supervised training using all the clean labels, (c) filters learned by the Rotation prediction, d) filters learned by solving the Jigsaw puzzle, (e) filters learned by solving the Jigmag puzzle, and (f) filters learned by SimCLR.}
\label{fig:histopathology_dataset_filters}
\end{figure}

In Fig. \ref{fig:covid_dataset_filters} and Fig. \ref{fig:histopathology_dataset_filters}, we compared the first layer filters learned by ResNet18 through various self-supervised tasks for the \textit{COVID-QU-Ex} and \textit{NCT-CRC-HE-100K} datasets. The results indicate that the filters learned by Rotation prediction and SimCLR are better aligned with the downstream task filters (Fig. \ref{fig:histopathology_dataset_filters}.b), especially in the \textit{NCT-CRC-HE-100K} dataset, compared to other pretext tasks. Despite the Jigsaw puzzle and Jigmag puzzle tasks learning more Gabor-like filters, as expected from supervised ImageNet training, initializing with such filters didn't yield substantial improvement in LNL methods compared to others. A plausible reason for this could be that the Jigsaw and Jigmag puzzle tasks train on image patches instead of the entire image, potentially resulting in features lacking global context, which is crucial for downstream tasks. In contrast, Rotation prediction and SimCLR models process the entire image, likely preserving the input image distribution closer to the original.

\end{document}